\documentclass[prb,a4paper,twocolumn,superscriptaddress,showpacs,amsfonts,amssymb,amsmath]{revtex4}
\usepackage{natbib}
\usepackage{graphicx}
\usepackage{dcolumn}
\usepackage{bm}

\begin{document}

\preprint{}

\newcommand{\LIO}{L1$_0$~}

    \title{Chemical order and size effects on the magnetic anisotropy of FePt and CoPt nanoparticles}
        \date{}

    \author{Stanislas Rohart}   \email{rohart@lps.u-psud.fr}
              \affiliation{Universit\'e de Lyon, F-69000, France; Univ. Lyon 1, Laboratoire PMCN; CNRS, UMR 5586; F69622 Villeurbanne Cedex}  
              \affiliation{Laboratoire de Physique des Solides, CNRS, Universit\'e Paris Sud, UMR 8502, B\^atiment~510, F-91405 Orsay cedex, France}
  \author{Florent Tournus}
              \affiliation{Universit\'e de Lyon, F-69000, France; Univ. Lyon 1, Laboratoire PMCN; CNRS, UMR 5586; F69622 Villeurbanne Cedex}
  \author{V\'eronique Dupuis} 
              \affiliation{Universit\'e de Lyon, F-69000, France; Univ. Lyon 1, Laboratoire PMCN; CNRS, UMR 5586; F69622 Villeurbanne Cedex}
  
    \begin{abstract}
       We investigate the consequence of the dimension reduction on the magnetic anisotropy of FePt and CoPt nanoparticles. Using an extension of the magnetic anisotropy model of N\'eel, we show that, due to a statistical finite size effect, chemically disordered clusters can display a magnetic anisotropy energy (MAE) as high as $0.5\times10^6$~J/m$^3$, more than one order of magnitude higher than the bulk MAE. Concerning \LIO ordered clusters, we show that the surface induces a reduction of the MAE as compared to the bulk, due to the symmetry breaking at the cluster surface, which modifies the chemical order.
    \end{abstract}

\maketitle

\section{Introduction}

The understanding and the control of the nanostructure magnetic properties is a challenging field of research, for both fundamental and applied physics. At the nanoscale, finite size effects and surface contributions become dominant and lead to a modification of the properties as compared to the bulk, such as the enhancement of the orbital magnetic moment and of the magnetic anisotropy energy (MAE) at the less coordinated atoms.\cite{gambardella2003,billas1994,rusponi2003,jamet2001} However, technological applications of such structures for ultrahigh density magnetic storage media are limited by thermal fluctuations of the magnetization direction. To overcome this so-called superparamagnetic limit, intense efforts are devoted to the investigation of materials with a high MAE. In such a view FePt\cite{sun2000,Dai2001,Rellinghaus2003,Antoniak2006,rong2006,rong2007} and CoPt~\cite{yu2002,xu2003,petit2004,Sun2004,mizuno2005,castaldi2005,tournus2008} have become some of the most studied materials as they display a very large
magnetocrystalline anisotropy (over $5\times10^6$~J/m$^3$) when ordered in the \LIO phase.

The effect of the \LIO ordering in FePt and CoPt on the magnetic anisotropy in bulk crystals or thin films has been widely studied on both experimental and theoretical points of view.\cite{Grange2000,razee2001,ersen2003,staunton2004,Skomski2007} In the case of nanostructures, as the surface modifies the local chemical environment, it is expected to have potentially a large impact on the MAE. Although many experimental studies had been performed\cite{sun2000,Dai2001,Rellinghaus2003,Antoniak2006,rong2006,rong2007,yu2002,xu2003,petit2004,Sun2004,mizuno2005,castaldi2005}, it is only recently that a clear impact of the dimension reduction on the MAE has been experimentally evidenced for CoPt nanoparticles.\cite{tournus2008} By using model samples, and combining various techniques, the authors have indeed been able to reliably determine the MAE of CoPt clusters with a diameter around 3~nm: while the anisotropy of chemically disordered particles was found to be higher than that of the bulk phase, it was the opposite for chemically ordered particles.

In this paper, we use an extension of the phenomenological anisotropy model of N\'eel to investigate the effect of chemical ordering and size reduction in both FePt and CoPt clusters. We first show that this simple model is in good agreement with bulk and thin film results. Then, taking advantage of its simplicity, which allows us to perform an extensive study of FePt and CoPt nanoparticles, we use this model to investigate the MAE of clusters. Taking into account finite size effects by mean of a statistical analysis on a large number of nanostructures, we demonstrate the impact of \LIO chemical ordering on the MAE and show how the surface and the finite size modify the results as compared to the bulk. Contrary to the results on pure Fe and Co clusters\cite{jamet2001,jamet2004}, here the surface induces a lowering of the MAE for \LIO ordered clusters.

Using a simple phenomenological pair-interaction model to calculate an effect as subtle as the magnetic anisotropy may seem a quite hazardous approach as compared to more elaborate models, like ab-initio calculations.
It is indeed well known that a theoretical determination of the MAE is a hard task: because it corresponds to very slight variations in the total energy as a function of the magnetization direction, highly converged results are required. Moreover, there is nowadays no available derivation of the N\'{e}el model from first principles. Nevertheless, this model, which is based on symmetries, has been found to be in good agreement with experiments on magnetic nanostructures\cite{jamet2001,jamet2004,morel2007}. In particular it can remarkably well 
account for the prevailing effect of low-coordinated atoms on the MAE. This let us think that this empirical model must be valid to a certain point, and that it can give reliable tendencies regarding the evolution of the MAE in nanostructures with various parameters (size, shape, configuration...). Besides, let us note that a N\'{e}el expression for the local anisotropy is also used in atomic scale magnetic simulations based on a Heisenberg-type model.\cite{kachkachi,jourdan2007}. Unfortunately, even it can be quite successfully compared to some ab-initio calculations and experiments (see further in the paper), we cannot \emph{demonstrate} the validity of the N\'{e}el model in the precise case under study: there is indeed no available ab-initio calculations of the MAE for comparable CoPt or FePt nanoparticles. However, in order to study the effect of size reduction and of chemical ordering in particles of a few nanometers (which is precisely the size of interest for ultra-high density magnetic storage applications), there is no other method that can reasonably be used, and that would be more precise. We are then convinced that investigating such complex systems, which are out of reach for first-principle calculations for the moment, will give valuable physical insights. Knowing the results predicted by this simple and crude model will then enable further comparison with more advanced calculations in the future. It is also important to check if the predictions of the N\'{e}el model are correct in the case of alloy nanoparticles, and we hope that our work will stimulate experimental studies.

\section{Magnetic anisotropy of a bimetallic alloy}

The magnetocrystalline anisotropy (MCA) arises from the coupling of the magnetization with the crystal lattice through the spin-orbit interaction\cite{bruno1989}. Then, for a given magnetic atom, it is directly linked to its local environment. For a whole structure, it is the signature of the system symmetries. It can be described using a phenomenological pair model, first introduced by L. N\'eel\cite{Neel1954}. In its original form, for a single element crystal, the MCA energy is written as a pair interaction over the first neighbors. The magnetic anisotropy
energy of a dimer is expressed, using the lowest order term as~\cite{Neel1954}
\begin{equation}\label{eq1}
    E_\mathrm{pair}=-L\left(\mathbf{m}\cdot\mathbf{e}\right)^2
\end{equation}

\noindent where $\mathbf{m}$ is the magnetization unit vector and $\mathbf{e}$ the unit vector along the bond. The prefactor $L$ is the anisotropy parameter, which has to be determined and is generally fitted on bulk experiments.

To consider a given structure, we need two more hypothesis. First we suppose that the magnetization direction is homogeneous over the whole structure (macrospin approximation), which is a quite suitable approximation for nanostructures with dimensions close to the exchange length.\cite{hubert1998} Second, we also suppose that the $L$ parameter is the same in the bulk and at the cluster surface, and can be extracted from bulk measurements.
Even if one could envisage a variation of the anisotropy parameter with the different environment in a structure, in fact 
the crude model of N\'eel with a uniform $L$ extracted from bulk measurements has already shown a good agreement (qualitative at least but also quantitative) for such systems.\cite{jamet2001,jamet2004,morel2007} As the purpose of our study is the prediction of the main tendencies, we will make this assumption in order to restrict the number of parameters in the model. On the basis of these two hypothesis, the total MCA energy, $E_\mathrm{a}$, is evaluated by performing a sum over all the nearest neighbor pairs. This leads to a quadratic form, that can be diagonalized into the orthonormal basis set $(\mathbf{e}_1,\mathbf{e}_2,\mathbf{e}_3)$. The normalization condition on  $\mathbf{m}$ further simplifies the expression to a biaxial form
\begin{equation}\label{eq:biax}
    E_\mathrm{a}/V=-K_1(\mathbf{m}\cdot\mathbf{e}_1)^2+K_2(\mathbf{m}\cdot\mathbf{e}_2)^2+K
\end{equation}

\noindent where $V$ is the volume of the structure, $K_1$ and $K_2$ are positive constants and $K$ a constant that can be omitted. Note that, in general cases, the new basis is different from the one defining the atomic lattice. The direction $\mathbf{e}_1$ (respectively $\mathbf{e}_2$) corresponds to the easy (respectively hard) magnetization direction, with the anisotropy constant $K_1$ (respectively $K_2$). This shows that this model, which is limited to second degree terms in
$\mathbf{m}$, always leads to a biaxial second order magnetic anisotropy displaying two degenerated energy minima. At the equilibrium (at zero temperature), the magnetization lies along $\mathbf{e}_1$ and $K_1V$ corresponds to the minimum energy needed to switch the magnetization from one minimum to the other one. Note that the value of $K_2$ does not influence this switching energy. For a bulk cubic crystal, this model would give a zero anisotropy,
as the second order expression is not compatible with the cubic symmetry (for a more accurate description of the MCA in this case, one should use higher order terms in eq.~\ref{eq1}, see ref. \onlinecite{morel2007}). However, this crude model is suitable to describe surface anisotropy where the cubic symmetry is locally broken.

To extend this model to an alloy containing two types of atom, named $A$ and $B$ in the following, three types of neighbor pair have to be considered:\cite{Skomski2007} two between identical atoms and one between $A$ and $B$ atoms. We introduce three different anisotropy parameters: $L_{AA}$, $L_{BB}$ and $L_{AB}$ respectively for $A$-$A$, $B$-$B$ and $A$-$B$ pairs. The MCA energy is then evaluated by summing eq.~\ref{eq1} over all the nearest neighbors pairs, taking into account the chemical nature of the two atoms.

This model allows us to describe the effect of \LIO chemical ordering on the MCA. Neglecting any tetragonal distortion, the atomic site lattice considered is face centered cubic (FCC), and we focus on equiatomic alloys. In this case, the \LIO phase corresponds to an alternation of pure $A$ and pure $B$ planes in the [001] direction, leading to a chemical anisotropy along the $c$~axis. Oppositely, in the chemically disordered phase named A1,
the atoms are randomly distributed on the lattice sites. The magnitude of the chemical order is usually quantified by the long-range order (LRO) parameter $S$, which varies between 0 (totally disordered alloy
) and 1 (perfectly ordered alloy
)\cite{Nix1938}. Defining two sublattices labelled $\alpha$ and $\beta$, corresponding to successive planes in the [001] direction, $S$ is given by 
\begin{equation}\label{eq:Sbulk}
S=2(n_\alpha-1/2)
\end{equation}

\noindent where $n_\alpha$ is the fraction of $\alpha$ sites occupied by atoms $A$. Note that changing the sign of $S$ is equivalent to inverting the convention for the choice of the $\alpha$ and $\beta$ planes. As this convention is arbitrary, the sign is meaningless and only the absolute value of $S$ has to be considered. Conversely, the probability $p$ for each lattice site to be occupied by an atom of the correct type ($A$ in a $\alpha$ plane and $B$ in a $\beta$ plane) is directly related to $S$ as $p=n_\alpha$.
%

To calculate the MCA of a bulk crystal of given order parameter, we consider the probability for a nearest neighbors pair to be a $A$-$A$, $B$-$B$ or $A$-$B$ bond and we calculate its average magnetic anisotropy parameter $\left\langle L\right\rangle$. Three different nearest neighbors pairs with equivalent $\left\langle L \right\rangle$ are obtained: those lying in the $\alpha$ planes, those lying in the $\beta$ planes and those linking atoms from two different planes. The corresponding $\left\langle L \right\rangle$ parameters are respectively $L_\alpha$, $L_\beta$, and $L_{\alpha\beta}$:
\begin{eqnarray}
  L_\alpha       &=& p^2 L_{AA}    + (1-p)^2 L_{BB}    + 2p(1-p)L_{AB}\nonumber\\
  L_\beta        &=& (1-p)^2L_{AA} + p^2 L_{BB}         + 2p(1-p)L_{AB}\nonumber\\
  L_{\alpha\beta}&=& p(1-p) L_{AA} + p(1-p) L_{BB} + [p^2+\left(1-p\right)^2]L_{AB}.\nonumber
\end{eqnarray}

For a chemically disordered crystal ($p=1/2$), the probability to have a $A$ or $B$ atom on a given site is identical, and $L_\alpha=L_\beta=L_{\alpha\beta}$: the cubic symmetry is recovered and the MCA is zero. As soon as a chemical order arises in the crystal ($S>0$), the cubic symmetry is broken and the second order MCA term has a finite value. Summing the different bond contributions in a FCC unit cell of volume $V_\mathrm{cell}$, the MAE is obtained as
\begin{eqnarray}
    E_{a,\mathrm{cell}}&=&2\left(L_\alpha+L_\beta-2L_{\alpha\beta}\right)   (\mathbf{m}\cdot\mathbf{e}_c)^2\nonumber\\
                       &=&2\left(L_{AA}  +L_{BB} -2L_{AB}         \right)S^2(\mathbf{m}\cdot\mathbf{e}_c)^2,
\end{eqnarray}

\noindent with $\mathbf{e}_c$ the unit vector aligned along the $c$ axis of the crystal. It corresponds to a uniaxial anisotropy, along the [001] direction ($c$ axis), and coincides with the chemical anisotropy direction. The anisotropy constant for the bulk alloy is then given by
\begin{equation}\label{eq:K(S)}
 K(S)=-2(L_{AA}+L_{BB}-2L_{AB})S^2/V_\mathrm{cell}
\end{equation}

\noindent and has a simple quadratic form. This evolution can be understood looking back at our model: as a pair interaction is considered, the model involves products of the occupation probabilities for a given site, which are linear in $S$.
If $2L_{AB}>L_{AA}+L_{BB}$, the anisotropy constant is positive, meaning that the $c$ axis is the easy magnetization axis with the anisotropy constant $K_1=K(S)$.

\begin{figure}[ht]
  \centering
    \includegraphics[width=0.9\columnwidth]{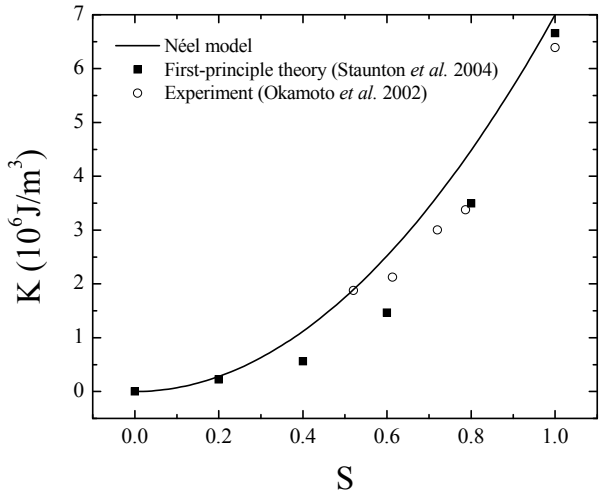}
    \caption{Variation of the anisotropy constant $K$ for a bulk FePt crystal versus the long-range chemical order parameter $S$. The results of our model (continuous line) are compared to experiments performed by S. Okamoto \textit{et al}. \cite{okamoto2002} and to first-principle calculations performed by J.B.~Staunton \textit{et al}. \cite{staunton2004} }
    \label{fig:DFT-Neel}
\end{figure}

We now apply this model to the case of XPt crystals, where $\mathrm{X}=\mathrm{Fe}$ or Co. In these alloys, a strong second order MCA accompanies the chemical ordering into the \LIO phase, and originates both from a local chemical environment change and from a tetragonal distortion of the FCC lattice. However, this slight deformation has a much weaker influence on the MCA\cite{staunton2004} and we neglect it in the following, as it was
supposed in the previous development. In order to apply our model, we need to determine the parameters $L_{\mathrm{XX}}$, $L_{\mathrm{PtPt}}$ and $L_{\mathrm{XPt}}$. As the magnetic moment on the Pt atoms is small in these crystals~\cite{Grange2000,Antoniak2006}, we neglect the magnetic
anisotropy of the Pt-Pt bonds by taking $L_{\mathrm{PtPt}}=0$. A straightforward relation is then obtained  between $L_{\mathrm{XX}}$ and $L_{\mathrm{XPt}}$ by fitting equation \ref{eq:K(S)} on the experimental MAE value of the \LIO ordered crystal. Consequently, there only remains one adjustable parameter and we finally estimate $L_{\mathrm{XX}}$ from measurements on pure samples using the original N\'eel model~\cite{jamet2004}:
taking into account the magnetostriction parameters, the model gives $L_{\mathrm{XX}}(r)=L_{\mathrm{XX}}(r_0)+(r-r_0)\frac{dL_{\mathrm{XX}}}{dr}(r_0)$,
where $r$ and $r_0$ are respectively the bond lengths in the alloy and in the pure crystal. We have determined the parameters for both FePt and CoPt (see table~\ref{tab:parameters}). The $L_{\mathrm{XX}}$ parameter for $\mathrm{X}=\mathrm{Fe}$ and \mbox{$\mathrm{X}=\mathrm{Co}$} has been calculated using the values given in ref.~\cite{jamet2004} and a bond length of respectively 2.72~\AA~for Fe and 2.71~\AA~for Co.
\begin{table}[ht]
    \centering\begin{ruledtabular}
        \begin{tabular}{lrrr}        
            XPt  & $K_{L1_0}$ & $L_{\mathrm{XX}}$ & $L_{\mathrm{XPt}}$ \\
                 & (J/m$^3$)               & (meV)             & (meV)              \\\hline
            FePt &                7.10$^6$ &             0.23  &            0.73    \\
            CoPt &                5.10$^6$ &             0.91  &            0.89    \\
        \end{tabular}\end{ruledtabular}
    \caption{Parameters $L_{\mathrm{XX}}$ and $L_{\mathrm{XPt}}$ of the N\'eel model for FePt and CoPt, determined as explained in the text from the                 \LIO ordered bulk MAE and the bulk Fe and Co N\'eel model parameters\cite{jamet2004}.}
    \label{tab:parameters}
\end{table}
In order to check the accuracy of our model, we compare some predictions with bulk and thin film results. The variation of bulk FePt MAE versus $S$ is compared to experiments\cite{okamoto2002} and first principle calculations\cite{staunton2004}, and, as it can be seen from fig.~\ref{fig:DFT-Neel}, shows a good agreement. The model is also compared to thin film experimental results. The MAE of a Pt/Co/Pt(111) multilayer with a Co thickness of one
monolayer is
\begin{equation}\label{eq:thinfilms}
    \frac {E_a}\Sigma=\frac{1}{s_{at}}\left(3L_{\mathrm{CoPt}}-\frac32L_{\mathrm{CoCo}}\right)=2K_S
\end{equation}

\noindent where $\Sigma$ is the film surface and $s_{at}^{-1}$ is the surface atomic density. $K_S$ is the usual surface anisotropy at the Co/Pt interface\cite{mcgee1993} (the factor $2$ on the right hand side of eq.~\ref{eq:thinfilms} accounts for the two Co/Pt interfaces in the film). Using the values of table~\ref{tab:parameters}, we find with our model $K_S=1.17$~mJ/m$^2$, close to the value of 1.15~mJ/m$^2$ reported in ref~\cite{mcgee1993}.

\section{Chemical order in nanoparticles}
\label{lab:secIII}

\begin{figure}[ht]
    \centering
    \includegraphics[width=0.9\columnwidth]{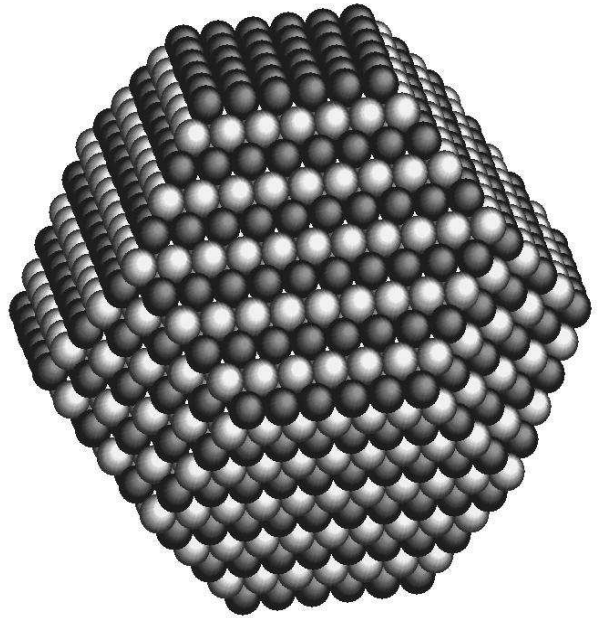}
    \caption{Sketch of a \LIO ordered truncated octahedron cluster containing 2406 atoms. The two colors represent the two types of atoms. Note that the topmost and bottommost planes contain the same kind of atom. Consequently, two types of cluster are possible: either the dark atoms correspond to A-type atoms, or they correspond to B-type atoms. These two clusters display different $f_\alpha$ factors, but have the same physical LRO parameter.}
\label{fig:cluster}
\end{figure}

In order to investigate the relationship between chemical order and magnetic anisotropy, we first need to precisely describe chemical order in nanostructures. Indeed, due to the finite size (and consequently the surface), the description is not as simple as in the bulk. First, the numbers of $\alpha$ and $\beta$ sites are not necessarily equal like in the bulk. The order parameter has to be adapted, so that a particle with a perfect alternation of A and B planes displays $S=1$. Using $f_\alpha$ the proportion of $\alpha$ sites, we use a more general definition of $S$:\cite{Nix1938}
\begin{equation}\label{eq:Smodif}
S=\frac{n_\alpha- x}{1-f_\alpha}
\end{equation}
\noindent where $n_\alpha$ is, as before, the fraction of $\alpha$ sites occupied by a A atom, and $x$ is the atomic concentration of A. Whereas for the bulk $f_\alpha=1/2$ and eq.~\ref{eq:Sbulk} is recovered, it varies depending on the shape and the size of a nanoparticle. Moreover, note that inverting the convention for $\alpha$ and $\beta$ planes changes $f_\alpha$ (for example, the truncated octahedron cluster, as the one represented in fig.~\ref{fig:cluster}, would display $f_\alpha=0.507$ or $0.493$ depending on the convention), but of course do not change the absolute value of $S$. Besides, a physical inversion of the atomic type on each lattice site will change $S$ into $-S$, meaning that the physical chemical order is unchanged. However, in this case, the inversion leads to a cluster with a different stacking (in particular, the surface atoms are not the same) and possibly with different magnetic properties, as it will be discussed later. 

To obtain a complementary description of the chemical order we can also use a short-range order (SRO) parameter, since for small size systems short-range ordering may dominate and control the nanoparticle properties. Although the SRO and LRO descriptions are equivalent for an infinite bulk (where there exists a direct relation between $p$, the probability for each site to be occupied by an atom of the correct type, SRO and LRO parameters\cite{cowley1950}) it is not true in nanoparticles.
As our anisotropy model is limited to the first neighbors, we also limit the SRO parameter to the nearest neighbors and use the description of Cowley.\cite{cowley1950} With $p_{l,m,n}$ the probability for an atom $i$ on a given site to have, in the $(l,m,n)$ direction, a nearest neighbor of the same kind, a local order parameter can be written as
\begin{equation}\label{eq:SRObulk}
    a_{l,m,n}=1-\frac{1-p_{l,m,n}}{x}\nonumber
\end{equation}

\noindent where $x$ is, as before, the atomic concentration in A. In an infinite AB FCC crystal and for a given atom, 12 of these parameters can be calculated. However, regarding the \LIO ordering, they are all equivalent (note that this is not true for all types of order): $a_{l,m,n}=a_i$ for $(l,m,n)$ directions in the $(a,b)$ plane and $a_{l,m,n}=-a_i$ otherwise, with $a_i=2p-1$ if atom $i$ is A-type on a $\alpha$ site or B-type on a $\beta$ site and $a_i=1-2p$ otherwise. One can then define the SRO parameter $\sigma$ as the average of the $a_i$ over the whole structure. Taking into account the occupation probability of each site, it leads to: $\sigma=(2p-1)^2$.

This parameter is directly linked to the LRO parameter as \mbox{$\sigma=S^2$}. Note that this quadratic relation is understood by the fact that the LRO description is based on the lattice sites occupation whereas the SRO description is based on nearest-neighbor pair correlations.

In a nanoparticle, the SRO parameter has to be adapted due to the surface, as it was the case for $S$. For each atom $i$, we calculate a local order parameter $\sigma_i$ as:
\begin{equation}\label{eq:sigma}
\sigma_i=\frac{n_i - x_i}{1-f_i}
\end{equation}
\noindent where $n_i$ is the proportion of nearest-neighbors (NN) of the same chemical type as atom $i$, in the $(\mathbf{a},\mathbf{b})$ plane, $x_i$ it the proportion of NN of the same chemical type as atom $i$, among all its NN, and $f_i$ is the proportion of NN in the $(\mathbf{a},\mathbf{b})$ plane among all the NN (if atom $i$ is surrounded by 12 NN like in the bulk, $f_i=1/3$; if it is at a surface site, $f_i$ may vary). The SRO parameter $\sigma$ is then defined as the average of $\sigma_i$ over the whole atoms. Of course, this definition of $\sigma$ is identical to the Cowley-type one for the infinite bulk.

An important feature, in the case of nanoparticles, is that the finite size allows us to explicitly take into account the statistical distribution of configurations. Indeed, the atomic arrangement for a given ordering probability $p$  is not unique, resulting in a statistical distribution of the chemical order parameters $S$ and $\sigma$. Then, the direct relation between $S$ and $p$ and the one between $\sigma$ and $S$, established for an infinite system, are no more valid. 

In order to investigate the relation between $p$, $\sigma$ and $S$ in nanoparticles, we have randomly generated a great number of nanoparticles with a given value of $p$. For each cluster, the true value of $S$ and $\sigma$ is calculated and we then obtain a statistical distribution of chemical order parameters (and of cluster composition), related to the small atom number in the cluster. Note that the choice of the direction defining the $\alpha$ and $\beta$ sublattices is completely arbitrary, so that the $[100]$ as well as the $[010]$ or the $[001]$ direction can be chosen to compute $S$ and $\sigma$. In order to have a physical description of the chemical anisotropy, we always choose the directions that gives, on one hand the highest value of $|S|$, and on the other hand, the highest value of $\sigma$.
For reasons explained further, we restrict our study to nanoparticles with a truncated octahedron shape.

\begin{figure}[ht]
    \centering
    \includegraphics[width=1\columnwidth]{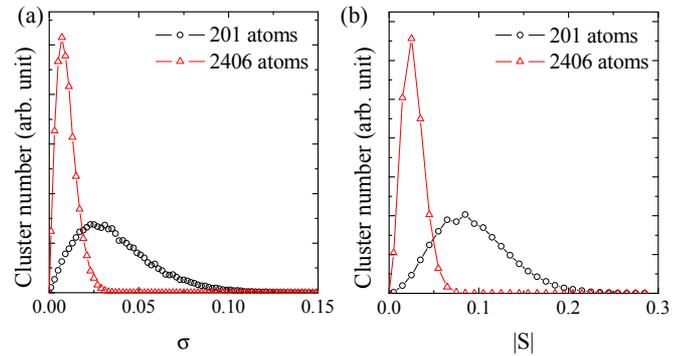}
    \caption{Distribution of the SRO ($\sigma$) and LRO ($|S|$) parameters for disordered
    clusters containing 201 atoms (circles) and 2406 atoms (triangles).}
    \label{fig:distribution_S}
\end{figure}

We first focus on the chemical order in clusters where each site has an equal probability to be occupied by a A or B atom (i.e. $p=1/2$). For each cluster, we have calculated both LRO and SRO parameters. Contrary to the bulk where a unique $S$ and $\sigma$ correspond to a given $p$, clusters display distributions of LRO and SRO, which are shown in fig.~\ref{fig:distribution_S}. As explained before, this is related to the small size of the particle and illustrates the distribution of configurations. This effect decreases when the diameter of the cluster increases, as the standard deviation of these parameters scales as the square root of the total number of atoms. Consequently, finite average values are found which means that the clusters cannot be considered as perfectly chemically disordered. The 201 atoms clusters display $\left\langle \sigma\right\rangle=0.037$ and $\left\langle |S|\right\rangle=0.094$ but note also that some clusters have a non-negligible probability to display chemical order parameter as high as $\sigma=0.08$ or $|S|=0.15$. Surprisingly, there is no detectable correlation between LRO and SRO for these clusters. This means that, for these disordered clusters, these parameters probe the chemical order in different and complementary ways and none of them is found to be better than the other one.

In partially ordered clusters ($p \neq 1/2$), the situation is quite different. Although a statistical distribution still exists, its width decreases with increasing $p$, which indicates that the number of configurations decreases. A clear correlation appear between LRO and SRO and their means follow the bulk relations: $\left\langle S \right\rangle\approx2p-1$ and $\left\langle\sigma \right\rangle\approx \left\langle S\right\rangle^2$.

In conclusion, we insist on the importance of the statistical distribution of configuration for nanoparticles.
Clusters generated with a given $p$ present a distribution of chemical order parameters, and conversely, a variety of configurations correspond to clusters with the same order parameter. This means that, for small particles, chemical order parameters are not sufficient to fully characterize the chemical arrangement of atoms in the cluster: it is impossible to sum up an entire chemical configuration with only one or two numerical parameters. A statistical analysis is then essential to correctly investigate the link between chemical order and the physical properties in nanoparticles.

\section{Magnetic anisotropy of nanoparticles}

We now use this anisotropy model to investigate the consequence of the size reduction in FePt and CoPt nanoparticles, with respect to their chemical order. We focus on nearly spherical clusters, as it corresponds to the usual shape experimentally observed. Moreover, as the host lattice for the FePt and CoPt crystal is FCC, we have chosen to restrict our study to clusters with a perfect truncated octahedron (TO) shape (see figure~\ref{fig:cluster}), which is the equilibrium shape in this case\cite{vanHardeveld1969,Dai2001,favre2006}. The magnetic anisotropy of pure Co TO clusters has already been studied using the N\'eel model\cite{jamet2004}. It has been shown that, due to the symmetries of the TO shape, both surface and volume MCA are zero (the MCA of the facets compensates) and that a surface anisotropy only arises when additional facets are present. Therefore, the MAE of perfect TO clusters, studied with our model
will explicitly reveal the effects of size reduction and chemical ordering. Moreover, for the sake of simplicity, and in order to avoid any effect coming from the variation of the magnetic element concentration, we have only considered clusters with a composition around 50~\% Pt. The dipolar interaction is also neglected in our study, since it is expected to play a significant role only for non-spherical clusters (shape anisotropy).\cite{rqdipolaire}

Concerning the magnetic anisotropy, whereas the bulk FePt or CoPt MCA is only uniaxial, the cluster finite size implies that this is no more true and the more general biaxial form (see eq.~\ref{eq:biax}) is always found. However, since the main point of interest is the minimum energy needed to switch the magnetization from an easy magnetization direction to the opposite one, we restrict our discussion to $K_1$.

\subsection{Chemically disordered nanoparticles}

\begin{figure}[ht]
    \centering
    \includegraphics[width=.9\columnwidth]{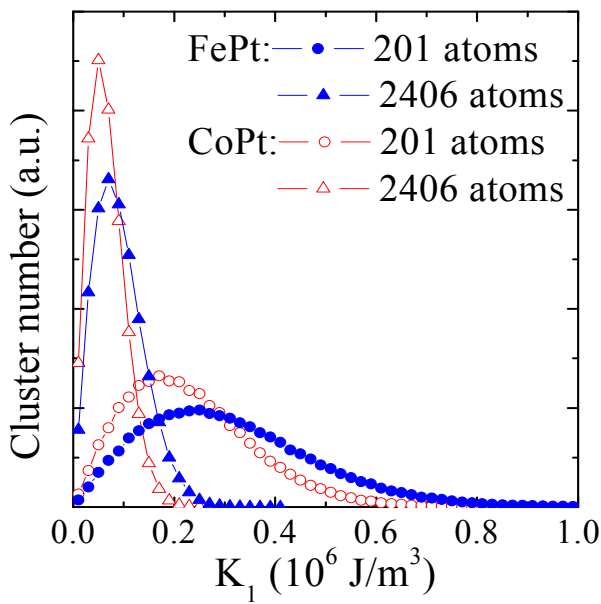}
    \caption{Distribution of  the magnetic
    anisotropy $K_1$ for FePt (close symbols) and CoPt (open symbols) disordered
    clusters containing 201 atoms - about 1.7~nm in diameter - (circles) and 2406 atoms  - about 4.3~nm in diameter - (triangles).}
    \label{fig:distribution}
\end{figure}

We first focus on the case of fully disordered clusters (\textit{i.e.} when $p=1/2$, each site having an equal probability to be occupied by a X or a Pt atom). For each cluster, we have calculated  $K_1$ by summing eq.~\ref{eq1} over all the nearest neighbors bonds. We obtain a statistical distribution which is displayed on figure~\ref{fig:distribution} for different types of clusters. Just like the $S$ and $\sigma$ distributions, the $K_1$ distribution is related to the small size of the cluster and its standard deviation scales as the square root of the total number of atoms in the cluster. Note also, that as for $|S|$ and $\sigma$, the mean value of $K_1$ decreases when the cluster size increases, indicating that for a large number of atoms the effect of
statistical inhomogeneities is smoothed out.

Besides, we have observed that the $|S|$ and $K_1$ distributions are statistically
independent which means that no detectable correlation exists between them.
At first sight, this may seem surprising. In fact these two quantities are very different:
$S$ probes the long-range chemical order whereas $K_1$ 
is sensitive to the local chemical order around the magnetic
atoms. A perfectly \LIO ordered bulk structure with an antiphase
boundary (\textit{i.e.} an inversion in the X and Pt planes
stacking sequence) exemplifies this difference: whereas $S$ will
be zero, the magnetic anisotropy will almost be unaffected. A
small $S$ is then not incompatible with a significant MCA. Note
however that a strong correlation exists between $|S|$ and $K_1$
when clusters display a higher chemical order, as it will be
discussed later. 

The local character of the magnetic anisotropy let us expect a 
stronger link between $K_1$ and the SRO parameter, because they both depend on NN correlations.
Indeed, we obtain a noticeable correlation between $K_1$ and $\sigma$. However, although the correlation coefficient between $K_1$ and $\sigma$ is significantly higher than the one between $K_1$ and $S$, and moreover increases with the particle size, it remains small. In particular, it corresponds to an insignificant variation of $\left\langle K_1\right\rangle$ with $\sigma$, as compared to the $K_1$ statistical distribution (this can be seen in figure~\ref{fig:K(sig)} for $\sigma \sim 0$). In small disordered particles, the configuration distribution is responsible for a partial ``decorrelation'' between the SRO and the magnetic anisotropy.
Although the SRO parameter is more suited than the LRO one, it still fails to
establish a definite link between chemical order and magnetic anisotropy.

\begin{figure}[ht]
    \centering
    \includegraphics[width=0.75\columnwidth]{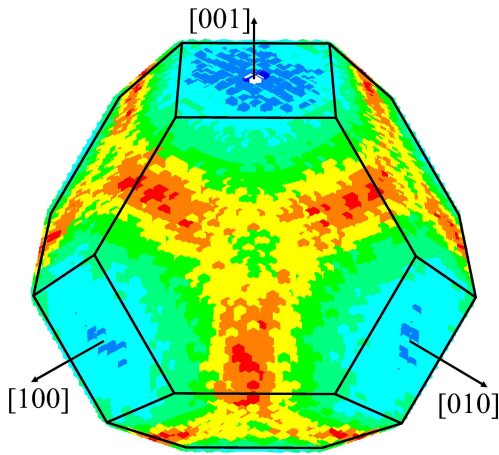}
    \caption{(color online) Statistical distribution of the easy magnetization axis for a 201 atom
    FePt cluster, represented in a 3D view along the [111] direction. The color scale
    represents the probability density for the easy magnetization axis to point in
    the corresponding direction, red (resp. blue) areas indicating the most (less)
    probable directions.}
    \label{fig:axisS0}
\end{figure}

We have also investigated the spatial distribution of the cluster
easy magnetization axis direction ($\mathbf{e}_1$ vector). Whereas
the \LIO ordering induces an anisotropy axis along the $c$ axis,
we do not find such a result for disordered clusters. The
$\mathbf{e}_1$ direction distribution satisfies the cluster cubic
symmetry and the three $a$, $b$ and $c$ axes are found to be
equivalent. However, the distribution is inhomogeneous and a
probability maximum occurs in the $[110]$ direction (and the other
equivalent directions), as shown in fig.~\ref{fig:axisS0}. This
means that the easy magnetization axis is rarely oriented towards
the \{001\} or \{111\} facets of the TO cluster but generally
points towards the edges connecting two \{111\} faces. Moreover,
this direction also corresponds to a maximum of the MAE: in the
case of the 201 atom FePt cluster presented in
fig.~\ref{fig:axisS0}, the mean MAE is about
$0.35\times10^6$~J/m$^3$ for an easy axis aligned along the [110]
direction while it is about $0.20$ and $0.25\times10^6$~J/m$^3$
respectively in the [001] and [111] directions.

\subsection{Chemical ordering}

\begin{figure}[ht]
    \centering
    \includegraphics[width=1\columnwidth]{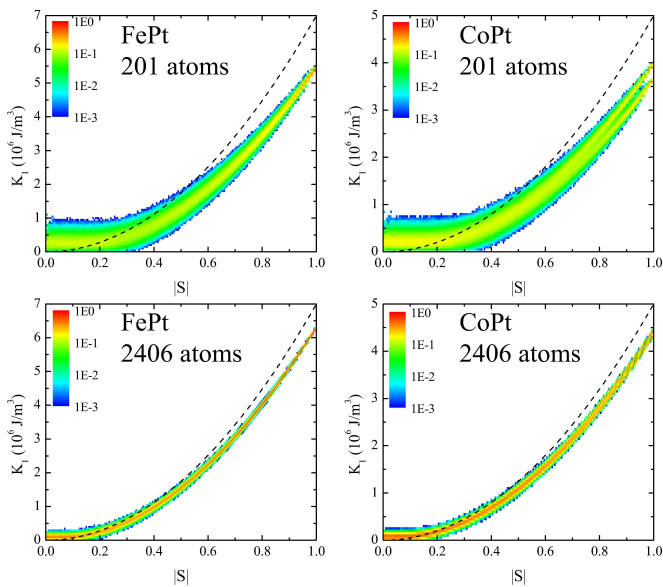}
    \caption{(color online) Variation of the anisotropy constant $K_1$ versus $|S|$ for a 201 atoms FePt
    (a) and CoPt (b) cluster, and for a 2406 atoms FePt (c) and CoPt (d) cluster. The color
    scale represents in the ($|S|$,$K_1$) plane, the density of probability for a cluster
    with chemical long range order parameter $|S|$ to have a MCA $K_1$ (the corresponding color bar
    is shown in inset). The dashed line corresponds to the bulk behavior.}
    \label{fig:K(S)}
\end{figure}

\begin{figure}[ht]
    \centering
    \includegraphics[width=1\columnwidth]{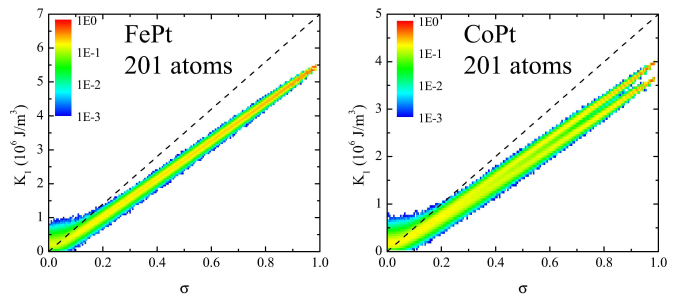}
    \caption{(color online) Variation of the anisotropy constant $K_1$ versus $\sigma$ for a 201 atoms FePt
    (a) and CoPt (b) cluster. The color
    scale represents in the ($\sigma$,$K_1$) plane, the density of probability for a cluster
    with chemical short range order parameter $|\sigma|$ to have a MCA $K_1$ (the corresponding color bar
    is shown in inset). The dashed line corresponds to the bulk behavior.}
    \label{fig:K(sig)}
\end{figure}

To study the chemical ordering effect in nanoparticles, clusters
with different $p$ values  
were generated, leading in each case
to a finite width distribution for $|S|$ and $\sigma$ (except for $p = 1$).
In this way, a large number of configurations is obtained allowing
us to investigate the relation between $|S|$, $\sigma$ and $K_1$.
The results are represented in fig.~\ref{fig:K(S)} and~\ref{fig:K(sig)} revealing 
the statistical nature of the link between $S$, $\sigma$ and $K_1$.
These Figures are analogous to the bulk $K(S)$ curve but also shows the finite width distributions.


Although in the bulk, simple relations have been found between $K_1$ and $S$ (quadratic variation) or $\sigma$ (linear variation), the relation is more complex for clusters. For small $|S|$ or $\sigma$, the mean $K_1$ is higher than the one of the bulk and we find the result previously described: versus $S$, the $K_1$
distribution is constant (mean value and width) and versus $\sigma$, a very small increase of the mean $K_1$ occurs. In this case, the
magnetic anisotropy is dominated by the small size effect, in
particular by the statistical distribution of the atoms on the
lattice sites. Above a threshold ($S_{th}$ or $\sigma_{th}$ respectively for LRO and SRO parameters), $\left\langle K_1 \right\rangle$
starts increasing and the distribution narrows (note that the threshold is the same for the two order parameters as $\sigma_{th}=S_{th}^2$). $\left\langle K_1
\right\rangle$ is proportional to $|S|^2$ or to $\sigma$, similarly to the bulk,
which means that the chemical order is sufficient to induce a
significant effect on the MAE. However, the increase is smaller
than for the bulk and the perfectly ordered \LIO cluster magnetic
anisotropy appears to be weaker than for the bulk. As it will be
discussed later, this is due to a surface effect. The case of CoPt
clusters is even more complex, as the magnetic anisotropy 
displays two distinct limits for perfectly ordered clusters. The difference is  $0.32 \times
10^6$~J/m$^3$ for the 201 atoms CoPt clusters. This corresponds to two
different \LIO ordered clusters where the Co and Pt atoms have been exchanged.
As previously discussed in section \ref{lab:secIII} this does not changes the order parameter but, while
in one case the top and bottom facets are made of Co atoms, in the other case
they are made of Pt atoms, the magnetic properties are different (the MAE is higher in this case).
This effect also exists
for FePt clusters but is weaker (the energy difference is
only $0.2 \times 10^4$~J/m$^3$) and it is not visible on the figures.
Finally, as the cluster size increases, the threshold decreases and the
$K_1$ value for \LIO ordered clusters increases, meaning that the behavior is getting closer to the bulk one.

\begin{figure}[ht]
    \centering
    \includegraphics[width=0.75\columnwidth]{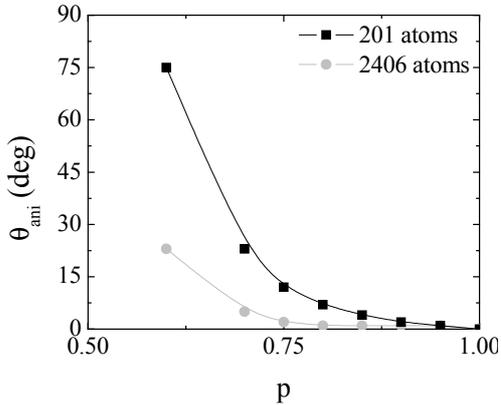}
    \caption{Variation of the anisotropy cone angle $\theta_{\mathrm{ani}}$
    (summit half angle) versus $p$ for 201 and 2406 FePt atoms clusters.
    The anisotropy cone is defined as the one, around the $c$ axis, containing
    90\% of the easy magnetization axes.}
    \label{fig:angle}
\end{figure}

Our calculations have also revealed that, as the chemical order increases,
the easy magnetization direction gets closer to the $c$ axis. 
When $p$ is high enough (which corresponds to $|S|$ above the threshold mentioned earlier),
the easy axis is contained in a cone around the $c$ axis whose summit
angle decreases as $p$ increases, as shown in figure~\ref{fig:angle}.

As a matter of fact, 
our results do not show any particular difference between the description using a SRO or a LRO parameter. In fact, as long as the chemical order is significant, a coherence is developed over the whole particle (statistically, a significant LRO goes with a significant SRO) and both parameters are highly correlated with the magnetic anisotropy. 
The SRO description is found to be slightly more convenient only 
for disordered clusters. On the other hand, the
theoretical variation of the magnetic anisotropy with the LRO parameter
may be more useful for comparisons with experiments, or with more advanced calculations,\cite{sun2000,razee2001,staunton2004,rong2006,rong2007} since the LRO parameter is usually used to quantify the degree of order in CoPt and FePt alloys. Moreover, while there is to our knowledge no
reported experimental measurement of a SRO parameter, for CoPt or FePt nanoparticles as those considered in our study, it is possible to determine their individual (using for instance transmission electron microscopy) or their mean LRO parameter.\cite{stappert2003,sato2005,petrova2005,saita2004}

\section{Discussion}

\begin{figure}[ht]
    \centering
    \includegraphics[width=0.9\columnwidth]{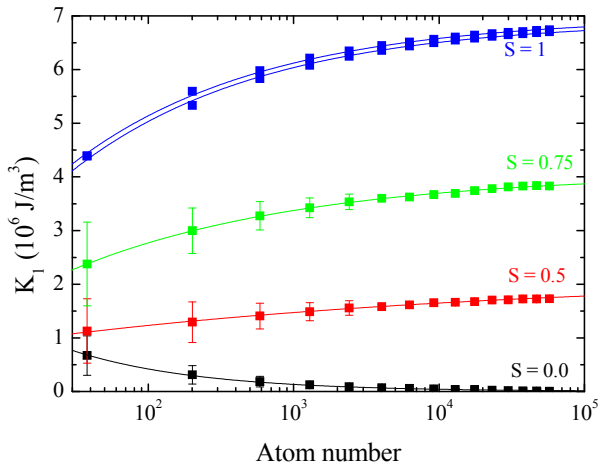}
    \caption{Variation of the mean anisotropy constant $\left\langle K_1 \right\rangle$
    versus the number of atoms in the cluster for FePt clusters of different $|S|$
    parameters. The dots correspond to the calculated $\left\langle K_1 \right\rangle$,
    the error bar representing the FWHM of the $K_1$ distributions. The lines are
    obtained by fitting the calculated values with the expression proposed in the text.}
    \label{fig:bilan}
\end{figure}

To apprehend the joint effects of surface and of small size in XPt
clusters ($\mathrm{X} = \mathrm{Co}$,Fe), we focus on two extreme
cases: fully disordered and perfectly \LIO ordered clusters.
Theses systems show a completely different variation of the MCA
versus the cluster size, as it can be seen in
figure~\ref{fig:bilan}. In the limit of large diameters, bulk
results are recovered: disordered clusters have a zero MCA and
\LIO ordered clusters display MCA of 7 and $5\times10^6$~J/m$^3$,
respectively for FePt and CoPt. For a finite diameter, chemically
disordered (resp. \LIO ordered) clusters display a higher (resp.
lower) MCA than the bulk one.

For perfectly ordered clusters ($S=1$), the variation of $\left\langle K_1 \right\rangle$
versus the number $N$ of atoms in a cluster can be adjusted by this expression 

\begin{equation}\label{eq:segal1}
    \left\langle K_1 \right\rangle(N)=K_V+\frac{K_a}{N^{1/3}}
\end{equation}

\noindent where $K_V$ is the bulk MAE, at the corresponding $S$,
and the second term is a small size correction.
This small size correction can be read in terms of a surface magnetic anisotropy because
it scales with the surface to volume ratio $\Sigma/V$. It could be alternatively written
as $K_S\Sigma/V$, with $K_S$ the surface magnetic anisotropy. Accordingly, the total anisotropy
energy for a particle can then be written, in an usual way, as the sum of a volume and
a surface contribution: $E_a = K_V V + K_S \Sigma$. However, in the present case $K_S$ is
negative, which is very unusual. The negative sign of $K_S$ indicates a lowering of the
MCA at the smallest sizes, as the chemical order is broken at the surface: due to missing
neighbors, the contribution of the XPt bonds is reduced and the total MCA decreases.
We find $K_S=-0.42$ and $-0.33$~mJ/m$^2$ for the FePt clusters, and $-0.32$ and
$-0.23$~mJ/m$^2$ for the CoPt clusters. For each type of clusters, two values for $K_a$
(and thus for $K_S$) are found due to the non-equivalence between the two \LIO stacking
possibilities, which indicates that the surface effects depend on the type of facets:
the clusters which have the highest anisotropy have pure Pt upper and lower $[001]$ facets.
This small size effect cannot be neglected for the smallest clusters: for less than 2000
atoms (about 3.5~nm in diameter), the decrease is more than 10~\% as compared to the bulk.
It is worth to mention that the absolute value of this surface magnetic anisotropy is
comparable to the highest one reported for pure clusters embedded in various matrices\cite{jamet2000,jamet2001b,rohart2006}.

For chemically disordered clusters ($S=0$) equation \ref{eq:segal1} cannot be used to account for 
the variation of $\left\langle K_1 \right\rangle$ with $N$. Interestingly, in this case the
evolution corresponds to the expression

\begin{equation}\label{eq:segal0}
    \left\langle K_1 \right\rangle(N)=K_V+\frac{K_b}{N^{1/2}}
\end{equation}

\noindent indicating that in this case, the small size contribution does not originate from
a surface effect. Here, the effect scales with the inverse square root of the number of atoms,
which is reminiscent of a statistical effect. Indeed, as discussed before, for small disordered clusters
the statistical distribution of the various chemical configurations is responsible for a concomitant
(but uncorrelated) enhancement of the mean long range order parameter and of the mean anisotropy constant.

In the case of intermediate $S$ values, the evolution of $\left\langle K_1 \right\rangle$ can 
be adjusted by a compination of the two preceeding expressions:

\begin{equation}\label{eq:sinter}
    \left\langle K_1 \right\rangle(N)=K_V+\frac{K_a}{N^{1/2}}+\frac{K_b}{N^{1/3}},
\end{equation}

\noindent where the two constants $K_a$ and $K_b$ respectively account for the
statistical/configuration effect and the surface effect.
The results of the fits are shown in table~\ref{tab:bilan}
for different values of $|S|$, both for FePt and CoPt clusters.
The relative weight of these two contributions gives us information on the origin
of the anisotropy variation with size. We find that the magnitude of the surface
contribution (i.e. the absolute value of $K_b$)
increases with the chemical order parameter of the particles, while the behavior
of the statistical contribution is less intuitive: $K_a$ is first increasing with $S$
and then finally vanishes when $S$ reaches 1. Note also that, since the surface and
the configuration contributions to the MAE have opposites signs, there exists a
range of $S$ values where there is almost no evolution of
$\left\langle K_1 \right\rangle$ with the particle size.

\begin{table}[ht]
    \centering\begin{ruledtabular}
        \begin{tabular}{lccccccc}

               & \multicolumn{3}{c}{FePt}    && \multicolumn{3}{c}{CoPt}    \\\cline{2-4}\cline{6-8}
            $|S|$  & \multicolumn{1}{l}{$K_V$}    & \multicolumn{1}{l}{$K_a$} & \multicolumn{1}{l}{$K_b$} && \multicolumn{1}{l}{$K_V$}    & \multicolumn{1}{l}{$K_a$} & \multicolumn{1}{l}{$K_b$} \\
\hline
            0    &  0.0     & 4.2     & 0.0  &&   0.0    & 3.0    & 0.0  \\
            0.5  &  1.9     & 6.4     &-5.9  &&   1.3    & 4.0    &-3.8  \\
            0.75 &  4.0     & 3.6     &-7.4  &&   2.8    & 3.1    &-5.0  \\
            1    &  7.0     & 0.0     &-9.2/-8.5  &&   5.0    & 0.0    &-6.9/-5.1 \\
        \end{tabular}\end{ruledtabular}
    \caption{Volume anisotropy $K_V$ and small size correction coefficients $K_a$ and $K_b$,
    determined for several $S$ using eq.~\ref{eq:sinter}, from the results in fig~\ref{fig:bilan}.
    The two different values of $K_b$ for \LIO ordered clusters account for the two types of
    \LIO stacking as discussed previously. Magnetic anisotropies are given in $10^6$~J/m$^3$.}
    \label{tab:bilan}
\end{table}

The predictions of the empirical N\'{e}el model can be of great help to analyze experimental
magnetic measurements. Indeed, according to the present results, we can expect the chemically
disordered particles to have a quite high anisotropy (much higher than that of the bulk phase),
whereas it should be in fact impossible to obtain small chemically ordered nanoparticles
having an intrinsic MCA as high as that of the bulk \LIO phase. This is consistent with our
very recent experimental findings on CoPt nanoparticles.\cite{tournus2008}
In addition, our calculations show that the anisotropy constant is falling off rapidly
when a particle is not fully ordered, which 
could also be responsible of the experimental observation of a lower MAE than expected.
This point may be
particularly relevant, since it is predicted that the most favorable configurations for small particles
correspond to $S<1$.\cite{muller2007} We emphasize that the results reported in this paper have been obtained
by completely letting aside any energetic consideration for the various chemical arrangements.
Nevertheless, our model could easily be combined with a structural optimization method,
either to determine the MAE of a few particular structures predicted to be the most stable ones,
or to determine the MAE distribution of a statistical ensemble of structures, taking into account
their probability of existence.

Another interesting feature predicted by our model is the significant MAE distribution,
even for a given particle size and a fixed chemical order parameter,
which is due to the existence of a statistical set of chemical configurations. As discussed
elsewhere,\cite{tournusieee} this can be one of the major source of MAE distribution in
realistic assemblies of clusters (with a distribution of composition, size, shape...). 
This feature, which is a particularity of magnetic alloys, should be directly visible
in experimental measurements on well-defined samples.

\section{Conclusion}

In this study, we have extended the anisotropy model of
N\'eel\cite{Neel1954} to describe the magnetocrystalline
anisotropy in mixed FePt and CoPt crystals. This phenomenological
model allows us to describe the effect of \LIO ordering on the MCA
and appears to be in good agreement with previously reported
results on bulk and thin films. We have used this model to
investigate the specific modifications occurring in nanosized
clusters of FePt and CoPt with a truncated octahedron shape. In
chemically disordered clusters, we have shown that there always
exists a partial chemical order, due to the small number of atoms,
and that, oppositely to the bulk, the MCA is non zero. 
On the contrary, for \LIO ordered clusters, we have shown that
surface induces a strong diminution of the magnetic anisotropy,
due to the broken \LIO periodicity at the cluster surface.

Our finding points a lack in this type of ordered alloys as materials for high-density magnetic storage applications, where a high switching energy $K_{\mathrm{eff}}V$ is needed to stabilize the magnetization direction. In pure or chemically disordered clusters, the anisotropy constant increases when the size decreases, thanks to a positive surface MAE (see e.g. ref.~\onlinecite{jamet2001,rusponi2003,rohart2006}), which attenuates the size variation of the switching energy. On the contrary, in the \LIO ordered CoPt and FePt clusters, $K_{\mathrm{eff}}$ decreases when the size decreases due to the negative surface MAE, which worthen the size variation of the switching energy. Here, surface is a drawback that may be compensated by playing on appropriate surface effects. Changing the surface anisotropy sign could be achieved using the high interface magnetic anisotropy that adds to surface magnetic anisotropy when the cluster surface is in contact with an appropriate non ferromagnetic material, like Pt\cite{jamet2001b} or CoO\cite{Skumryev2003}. Surface engineering, by embedding the clusters in the matrix or by synthesizing core-shell clusters with the non ferromagnetic shell around the magnetic core, appears to be complementary approaches to these \LIO ordered materials, in order to overcome the superparamagnetic limit in nanoscale clusters.

\begin{acknowledgments}

We gratefully acknowledge V. Repain and Y. Nahas (MPQ, Paris, France) for
stimulating discussions. We acknowledge support from the European
Community (STREP SFINx no. NMP2-CT-2003-505587).

\end{acknowledgments}


\begin{thebibliography}{50}
\bibitem[{\citenamefont{Note}(2008)}]{rqdipolaire}
\bibinfo{author}{\bibfnamefont{We have checked this point by explicitly calculating the shape anisotropy. Indeed, because we are dealing with alloys, even the highly symmetric truncated octahedron shape does not give zero dipolar anisotropy (shape anisotropy), but according to our calculations it remains negligible}}.
\end{thebibliography}
\end{document}